\documentclass[journal]{IEEEtran}
\newcommand{\PreserveBackslash}[1]{\let\temp=\\#1\let\\=\temp}

\usepackage{color}
\usepackage{array}
\usepackage{multirow}
\usepackage{textcomp}  
\usepackage{cite}
\usepackage{graphicx}
\usepackage{cite}
\usepackage{subfigure}
\usepackage[T1]{fontenc}

\graphicspath{{./fig/}}

\begin{document}
\title{Framework for High-performance Video Acquisition and Processing in MTCA.4 Form Factor}

\author{
	A.~Mielczarek, D.~Makowski,~\IEEEmembership{Member,~IEEE}, P.~Perek, A.~Napieralski,~\IEEEmembership{Senior Member,~IEEE}
	
	\thanks{A.~Mielczarek, D.~Makowski, P.~Perek, A.~Napieralski,
			are with the Lodz University of Technology, Poland (e-mail: amielczarek@dmcs.pl)}

}

\maketitle

\section{Introduction}

The video acquisition and processing systems are commonly used in industrial and
scientific applications. Many of them utilize Camera Link interface for the
transmission of a video stream from the camera to the host system. The framework 
presented in the paper enables capturing such data, processing it and transmitting
to the host CPU. It consist of MTCA.4-compliant frame grabber and a set of software
libraries supporting several different cameras. It is designed for use in large scale
physics experiments such as ITER tokamak or European X-Ray Free-Electron Laser (E-XFEL),
as well as in the Centre for Free-Electron Laser Science (CFEL).

\section{MTCA.4}

The base MTCA.0 specification allows building high-throughput
telecommunication solutions, but is not ideally suited for Data Acquisition (DAQ) systems.
Although the MCH usually allows advanced routing of the clock signals, the infrastructure for
distributing synchronization and trigger signals is limited. Additionally, there is no possibility of
connecting the cables from the rear side of the shelf -- which is a recommended practice in many
applications. All the cables have to enter the modules through the front
panel often reducing accessibility of the neighboring modules~\cite{MTCA_0}.

\begin{figure}[htb]
\centering
\includegraphics[width=0.4\textwidth]{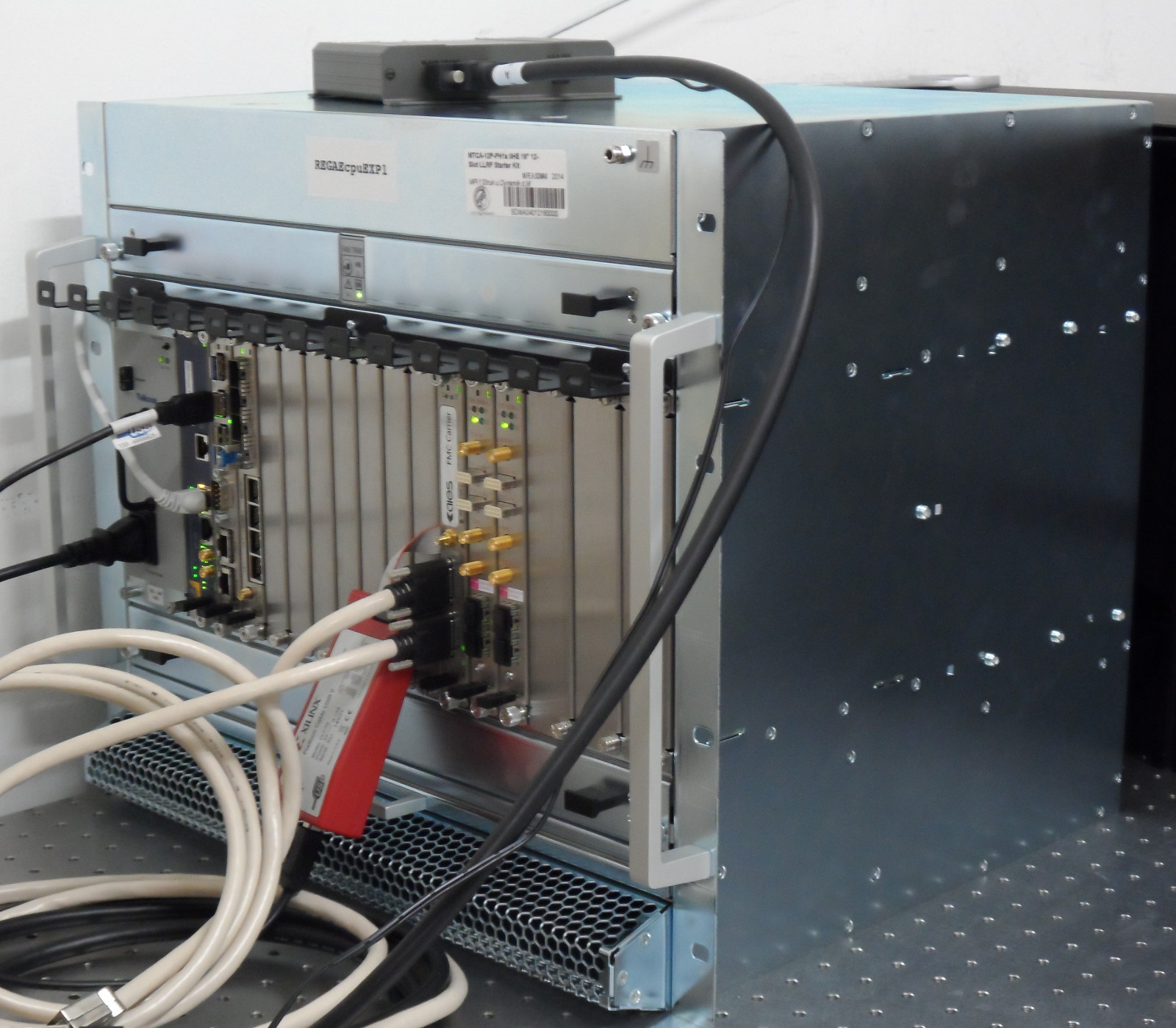}
\caption{A typical MTCA.4 12-slot chassis}
\label{fig:mtca4_shelf}
\end{figure}

All the aforementioned issues were addressed in the MTCA.4 subsidiary specification "MicroTCA
enhancement for rear I/O and precision timing". A typical MTCA.4-compliant crate is presented
in Figure~\ref{fig:mtca4_shelf}.

The MTCA.4 specification defines an M-LVDS bus connecting all the AMC slots which allows easy
distribution of system-wide signals, like timing events or interlocks. It also introduces concept
of a Rear Transition Module (RTM) -- a second PCB closely cooperating with the AMC module. Such
a module not only effectively doubles the available PCB space but also allows connecting the
cables in the rear of the shelf~\cite{MTCA_4}.

\section{Hardware Platform}

The commercially available MFG4 solution was selected as target platform for the frame grabber
infrastructure reference implementation. It is composed of a dual slot FMC carrier board and
of Camera Link pass-through modules. The carrier card is a double width AMC module with a recent
\mbox{7-Series} Xilinx Artix FPGA. Depending on the option, it is accompanied with one or two
FMC Camera Link modules. The deserialization is meant to be done directly in the FPGA,
thus no external active circuits are needed.
The MFG4 base board is equipped with a 2~GB SDRAM buffer consisting of four 16-bit DDR3 chips. The
64-bit DDR bus can operate at 533~MHz offering a throughput of 68.2~Gb/s. This is more than twice
the throughput required to support two cameras outputting video at the maximum possible data rate
supported by the Camera Link standard (full configuration, 80-bit mode).

\section{The FPGA Framework}

Although the Camera Link standard was first released in 2000, it is still commonly used in
high-speed cameras. The most efficient 80-bit mode of operation allows 80 bits of data plus 4
synchronization signals to be transmitted in every clock cycle. The clock frequency is limited by
the standard to 85~MHz. The data to be transmitted is split into three groups of 28 bits and serialized.
Each group is transmitted over 4 data lines accompanied by a clock signal, following the
Channel Link protocol.

Receiving a Camera Link data is not trivial, as the link is composed of three independent, possibly
asynchronous, channels. Therefore, each channel has to be buffered in the receiver and then synchronized
with the others. The Camera Link standard defines a variety of mappings between pixel data and link words.
Rearrangement of the incoming data is partly done in firmware and partly in software due to a large number
of possible configurations.

The video data is captured by the Camera Link receiver and transferred via an AXI Stream interface.
The video stream is then directed to the Xilinx Virtual FIFO (VFIFO). It is an IP-Core which helps
implement FIFO queues based on external memories. The virtual FIFO is a very helpful component,
however it also has some drawbacks. For instance, it can only control relatively small memory regions
(up to several tens of MB). Moreover, it cannot emit a warning when the amount of available storage
space drops below a certain limit. Therefore, a dedicated monitoring component was developed to
asses if the next frame will fit in the memory or not.

During write of the captured data to the VFIFO the frame resolution is identified and a corresponding
timestamp is captured. This information, along with user supplied meta-data, are stored into a
frame information FIFO. The synchronization signals form the camera are not stored in the VFIFO.
These are however, crucial for further data processing. Moreover, the processing algorithms should
have the information on the image resolution in advance. To accommodate for these features, the frame
reader module was developed. Firstly, it reads the information on frame resolution and then reads the
actual frame data from the VFIFO. It is able to fetch just the right amount of data from the memory
and regenerate the synchronization signals. With each retrieved frame the module provides information
on its resolution, number, timestamp, etc.

\begin{figure}[htb]
\centering
\includegraphics[width=0.4\textwidth]{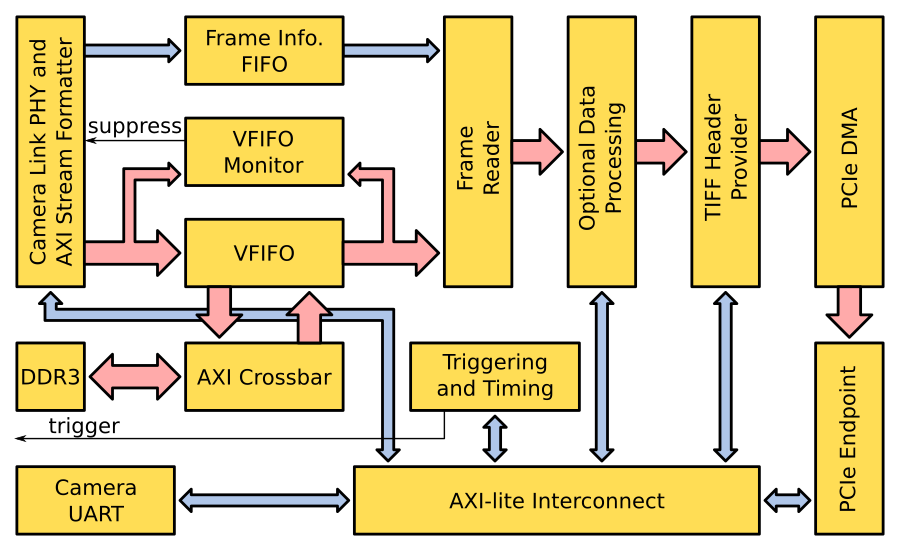}
\caption{Block diagram of the FPGA firmware}
\label{fig:firmware_top}
\end{figure}

Before the data is stored in the host memory, each captured and processed frame has to be appended
with a header. The headers are generated by the TIFF header provider. The TIFF format was
selected mainly due to its straightforward file structure and support for uncompressed gray-scale images.

The video stream is written to the host computer memory by a custom DMA engine developed at
the Department of Microelectronics and Computer Science (DMCS) of Lodz University of Technology.
The engine supports efficient transfers utilizing a scatter-gather list.
The camera is controlled and monitored through a simple UART interface. The physical layer is defined
by the Camera Link specification, whereas the frame format and particular command set is vendor dependent. 

\section{Software Support}

The framework also includes complete software support consisting of Linux device driver, API libraries as well as console and graphical user applications. The Linux device driver implements two independent interfaces: one for high-performance image acquisition and the other for camera control using UART interface defined in the Camera Link standard. The driver fully supports the DMA engine implemented in the FPGA firmware and ensures efficient data transfer from the FPGA buffer directly to user-space memory~\cite{perek2013high}. 
To facilitate software development the framework also provides set of API libraries. The low-level library communicates directly with the device driver and provides easy-to-use functions for image data reading and sending/receiving characters via the UART interface. The higher-level libraries implement control protocols defined for particular cameras. Usually single library is dedicated for specific camera or group of cameras sharing the same control protocol. Thanks to such approach, the framework can be easily extended to support new cameras. The only effort necessary for communication with new device is development of camera library implementing its control protocol.
The highest layer of the software support are console and graphical applications. All of them use camera-specific libraries for communication with the hardware. The console applications provide easy access to current settings and status information, control of the camera and image acquisition. The graphical applications additionally offer data visualization (both live preview and off-line data display) and basic processing. Graphical user interface allows convenient camera configuration, data acquisition, recording and analysis.

\section{Proof of Concept System}

The developed framework enables collecting the data from the camera using the top performance
80-bit mode of the Full Camera Link interface, offering 6.8~Gb/s of raw image data throughput.
The solution was tested with a number of commercially available cameras, including:
Mikrotron MC3010/MC3011, PCO EDGE 5.5, Andor NEO 5.5, Basler Sprint spL2048-70km.

\section{Conclusion}

The proposed video acquisition solution features the worlds first Camera Link frame grabber for
the MTCA.4 architecture. MFG4 card was chosen due to its fair cost to performance balance
and provision of a Xilinx \mbox{7-Series} FPGA. Thanks to the modern FPGA circuit architecture,
the deserialization is done using only the built-in ISERDES primitives, which reduces the costs
and complexity of the required hardware. The implemented deserializer is capable of supporting
all the Camera Link modes and configurations. Especially, it allows capturing the Full Camera
Link 80-bit image data with reference clock of 85 MHz, which constitutes the highest data rate
supported by the standard.

\bibliographystyle{IEEEtran}
\bibliography{biblio}

\end{document}